\documentclass[prd,aps,
twocolumn,
nofootinbib,
floatfix,
superscriptaddress,
preprintnumbers]{revtex4}
\usepackage{graphicx,epstopdf}
\usepackage{epsfig}
\usepackage{rotating}
\usepackage{amssymb}
\usepackage{subfigure}
\usepackage{dsfont}
\usepackage{psfrag}
\usepackage{amsmath,euscript,array,mathrsfs}
\usepackage{axodraw}
\usepackage{bbold}

\newcommand{\beq}{\begin{equation}}
\newcommand{\eeq}{\end{equation}}
\newcommand{\beqs}{\begin{eqnarray}}
\newcommand{\eeqs}{\end{eqnarray}}
\newcommand{\lsim}{\mathrel{\raisebox{-
.6ex}{$\stackrel{\textstyle<}{\sim}$}}}
\newcommand{\gsim}{\mathrel{\raisebox{-
.6ex}{$\stackrel{\textstyle>}{\sim}$}}}

\def\hbar{\hspace{0pt}\raisebox{1pt}{$-$} \hspace{-7pt} h}

\def\di{\mbox{d}}
\def\r{\rho}

\newcommand{\be}{\begin{equation}}
\newcommand{\ee}{\end{equation}}
\newcommand{\bea}{\begin{eqnarray}}
\newcommand{\eea}{\end{eqnarray}}

\def\lbldef#1#2{\expandafter\gdef\csname #1\endcsname {#2}}

\def\href#1#2{#2}

\newcommand{\ber}{\begin{eqnarray}}
\newcommand{\eer}{\end{eqnarray}}

\newcommand{\beqar}{\begin{eqnarray}}

\newcommand{\eeqar}{\end{eqnarray}}

\newcommand{\dsl}
  {\kern.06em\hbox{\raise.15ex\hbox{$/$}\kern-.56em\hbox{$\partial$}}}

\newcommand{\eeqarr}{\end{eqnarray}}
\newcommand{\ZZ}{{\rm \kern 0.275em Z \kern -0.92em Z}\;}

\def\CC{{\mathchoice
{\rm C\mkern-8mu\vrule height1.45ex depth-.05ex
width.05em\mkern9mu\kern-.05em}
{\rm C\mkern-8mu\vrule height1.45ex depth-.05ex
width.05em\mkern9mu\kern-.05em}
{\rm C\mkern-8mu\vrule height1ex depth-.07ex
width.035em\mkern9mu\kern-.035em}
{\rm C\mkern-8mu\vrule height.65ex depth-.1ex
width.025em\mkern8mu\kern-.025em}}}

\def\RR{{\rm I\kern-1.6pt {\rm R}}}

\def\ZZ{{\rm Z}\kern-3.8pt {\rm Z} \kern2pt}
\def\IB{\relax{\rm I\kern-.18em B}}
\def\ID{\relax{\rm I\kern-.18em D}}
\def\II{\relax{\rm I\kern-.18em I}}
\def\IP{\relax{\rm I\kern-.18em P}}

\newcommand{\bear}{\begin{eqnarray}}
\newcommand{\eear}{\end{eqnarray}}

\def\to{\rightarrow}

\def\to{\rightarrow}

\def\r{\rho}                                     
\def\t{\tau}

\def\6{\partial}

\def\bea{\begin{eqnarray}}
\def\eea{\end{eqnarray}}

\def\beqx{\begin{displaymath}}
\def\eeqx{\end{displaymath}}

\newcommand{\bmat}{\left(\begin{array}}
\newcommand{\emat}{\end{array}\right)}

\def\r{\rho}

\def\t{\tau}

\def\bo{{\raise-.3ex\hbox{\large$\Box$}}}               
\def\face{{\raise.2ex\hbox{$\displaystyle \bigodot$}\mskip-2.2mu \llap {$\ddot
        \smile$}}}                                   
\def\>{\rangle}                                      
\def\<{\langle}                                      

\def\leftrightarrowfill{$\mathsurround=0pt \mathord\leftarrow \mkern-6mu
        \cleaders\hbox{$\mkern-2mu \mathord- \mkern-2mu$}\hfill
        \mkern-6mu \mathord\rightarrow$}        
\def\dvec#1{\vbox{\ialign{##\crcr
        \leftrightarrowfill\crcr\noalign{\kern-1pt\nointerlineskip}
        $\hfil\displaystyle{#1}\hfil$\crcr}}}           

\def\-{\hphantom{-}}

\begin{document}

\title{Gauge/gravity dualities and bulk phase transitions.}

\author{Anton F. Faedo}
\affiliation{Departament de F\'{\i}sica Fonamental \&  Institut de Ci\`encies del Cosmos, Universitat de Barcelona, Mart\'{\i}  i Franqu\`es 1, E-08028 Barcelona, Spain.}
\author{Maurizio Piai}
\affiliation{Department of Physics, College of Science, Swansea University,
Singleton Park, Swansea, Wales, UK.}
\author{Daniel Schofield}
\affiliation{Department of Physics, College of Science, Swansea University,
Singleton Park, Swansea, Wales, UK.}

\date{\today}


\begin{abstract}
\noindent We consider D7-branes probing several classes of Type IIB supergravity backgrounds,
and study the classical problem of finding equilibrium configurations for the 
embedding functions. This is a method employed to model chiral symmetry breaking
in the gravity dual of a strongly-coupled, confining gauge theory.
We unveil and discuss a new type of phase transition appearing in
the gravity systems, which is  similar in nature and meaning to bulk phase transitions on the lattice.
The existence of this genre of phase transition puts a new, intrinsic limit
on the region of parameter space which can be used to study the physics of the
dual field theory. We complete the analysis of  D7 embeddings in wrapped-D5
 supergravity backgrounds, and explain in what cases chiral-symmetry breaking 
 is sensibly modelled by the gravity construction.
 
\end{abstract}

\preprint{ICCUB-14-043}

\maketitle

\tableofcontents

\section{Introduction}

The study of strongly-coupled field theories 
is notoriously difficult. Traditional perturbation theory  fails to capture
non-perturbative phenomena, 
that become dominant at strong coupling.
In the case of (four-dimensional) gauge theories
there are two examples in which strong dynamics needs to be address:
the study of QCD 
and
of technicolor~\cite{TC,reviewsTC,WTC},
a proposal for the origin of electroweak symmetry breaking (EWSB).
In both cases, the strong dynamics induces
 the spontaneous breaking of (approximate) 
chiral symmetry.

A powerful tool in the study of realistic gauge theories in the strong-coupling regime 
is lattice field theory: one replaces the continuum space-time with a discretized space,
which is amenable to numerical studies. 
Besides its practical  utility, the lattice provides a sensible
definition of the measure in the path integral.

String theory on curved backgrounds
offers an alternative way of studying four-dimensional, strongly-coupled,
field theories~\cite{AdSCFT,reviewAdSCFT}. This framework goes under the generic 
name of {\it gauge/gravity dualities},
and has the peculiarity that when the field theory is at strong coupling, its gravity dual (under suitable conditions)
is weakly coupled.
 For example, this approach allows one to compute the Wilson loop of a gauge theory~\cite{MaldacenaWilson},
to characterise confinement~\cite{Witten} and to discuss the properties of matter
 fields in a gauge theory~\cite{KK}.
  
Gauge/gravity dualities are based on a conjecture. Aside from strong
evidence collected by systematically comparing physical 
quantities, certain formal procedures that are
necessary in field-theory calculations
can be seen to emerge also in the treatment of their gravity duals.
The obvious example 
is holographic renormalisation~\cite{HR}:
 when extracting physical (field-theory) results for correlation functions from the gravity theory, one
has to implement a procedure closely reminiscent of renormalisation 
in field theory.

In this paper we show an 
 example of a formal problem,
emerging in certain gravity calculations, 
 that has an analog in lattice gauge theory, and that forces us to 
 implement a procedure similar to what is done 
 on the lattice. On the one hand this provides another, new element 
 to support the basic ideas of gauge/gravity dualities. On the other hand 
  it  has unexpected and important physical consequences, placing intrinsic limits on the regime of applicability of the dualities.
 
 In the lattice literature, the analog of the phenomenon we want to discuss is that of {\it bulk} phase transitions.
 Generically, a lattice theory is different from the continuum field theory it intends to model.
  The finite lattice spacing  
  introduces a new scale that can be thought of as some 
 UV-regulator for the field theory. Furthermore, the operators
 (and symmetries) of the lattice 
 are different from those of the continuum theory.
 However, provided the long-distance physics
of the lattice and field theories are in the same universality class,
  the lattice calculation yields useful results that also hold true in the continuum.
  The field theory is defined in terms of an RG fixed point of the lattice theory.
 
 But it may occur that the lattice theory has
 a non-trivial phase structure, with lines (or hypersurfaces)
  of phase transitions (or crossovers) partitioning the parameter space.
  Thus, one has to make sure that the 
  region of parameter 
  space that the calculation is probing 
  is restricted to the patch 
  that is connected to the relevant fixed point,
  and hence belongs to the same universality class as the field theory.
  In general, this imposes bounds on the magnitude of the parameters in the lattice action.

  An example of this phenomenon is the bulk phase-transition that appears in 
  $SU(N)$ Yang--Mills
   theories  as a function of the lattice coupling $\beta\sim\frac{1}{g^2}$. One can perform 
  two systematic expansions of the lattice action, at small-$\beta$ (strong coupling) and large-$\beta$ 
  (weak coupling), and the results from the two cases do not connect smoothly.
More sophisticated arguments reveal that there is a critical $\beta_c$ at which a first-order
phase transition (or a crossover) takes place. The large-$\beta$ regime is smoothly connected
to the $g\rightarrow 0$ limit, and hence to the original field theory.
On the contrary, the small-$\beta$ region is not. 
From the point of view of the lattice, the theory defined by small choices of $\beta$ 
is  perfectly well-defined
and might be worth studying per se: 
 the strong-coupling 
expansion yields 
the area-law of confinement for the Wilson loop.
Unfortunately, from the field-theory perspective this is purely a lattice artifact: all the 
physics is governed by the scale set by the lattice, and not by the Yang--Mills  dynamics.
Another interesting example of bulk phase transition, 
in which the phase-space structure of the lattice theory is 
more complicated, has been reconsidered recently in the 
context of $SU(2)$ pure gauge theory with a mixed fundamental-adjoint action for the gauge fields~\cite{LPRR}.
 
We find a  similar situation in gauge/gravity dualities.
 We will perform a set of classical gravity calculations in the presence of a 
boundary that is needed as a UV regulator. 
The physics will depend on the dynamics represented 
 by the curved background, but also on some boundary-valued {\it control parameter} which 
 is assumed to have a field-theory origin. 
 The calculations can be done for all possible
 values of the control parameter, and produce physically sensible results in gravity.
 However, in special cases there are phase transitions such
 that only on one side of the phase space it is possible to remove the
 unphysical cutoff, and recover the dual field theory.
 On the wrong side of the transition, by contrast, observables 
 are dominated by the 
 cutoff scale. Ultimately this yields new, unexpected bounds on the
region of parameter space in gravity
which is related to the dual field theory.

We do not develop a systematic study of this feature.
We focus on a specific framework and set of problems 
that are physically interesting, observe under what conditions
these bulk transitions appear, and draw  conclusions on
the physics relevant to the dual field theories. 

We consider a class of Type IIB (super-)gravity solutions,
generated by a large number of D5-branes wrapping a two-cycle within the base of the conifold~\cite{CO},
which includes the CVMN  model~\cite{MN,CV}.
Several subclasses of backgrounds
of this type exhibit multi-scale dynamics,
and have been used as a starting point 
to build the dual of  technicolor theories~\cite{NPP} which admit a light
composite scalar in the spectrum~\cite{ENP,EP} (see also~\cite{Elander:2014ola}).

Chiral symmetry breaking has been modelled 
and studied on some of these  
backgrounds~\cite{A,ASW1,ASW2,CLV,ASW3,ASW4}
by probing them with D7-branes with a special embedding,
in analogy with what was done in~\cite{SS} and~\cite{KuSo,DKS}.
The interest in this set-up arises 
 because the multi-scale dynamics might 
help in resolving the phenomenological problems
 (in particular with the  $S$ parameter) 
of models of EWSB obtained within other frameworks~\cite{stringS}.

This paper complements the work presented in~\cite{FPS} by completing the analysis to include backgrounds 
that were not considered there.
The two together cover the entirety of supersymmetric 
wrapped-D5 solutions that 
do not show bad singular behaviours. With the purpose of illustrating what bulk transitions are, 
we also study a different class of Type IIB backgrounds
(the flavoured Abelian solutions), which have the main advantage of being
peculiarly simple.

The paper is organised as follows.
In Section~\ref{Sec:gencons} we introduce the general formalism needed in the rest of the analysis.
In Section~\ref{Sec:wrappedD5} we provide the reader with all the useful details about the supergravity solutions. 
In Section~\ref{Sec:D7embedds} we compute the D7 
embeddings, and explicate what bulk transitions are by means of an accessible example.
In Section~\ref{Sec:chiralbreak} we consider the remaining backgrounds in the wrapped-D5 class, and provide a summary of all the results of this paper together with~\cite{FPS}.
We conclude in Section~\ref{Sec:conclusions} by critically discussing our findings and suggesting directions for 
further studies.

\section{Discussion.}
\label{Sec:gencons}

In this section, we summarise 
known general results and fix the notation, keeping explicitly the 
dependence on the UV regulator. 
Some of the considerations developed here can be found, in different but
related contexts, in~\cite{BS, ASS, GO, BKY,  NPR,  APT,  FPS}.
We then discuss the case in which a certain class of
phase transitions occur in the presence of a finite cutoff, which 
would be missed by a too naive implementation of the renormalisation procedure.

Let us consider the classical system with the action:
\beqs\label{genaction}
{\cal S} &=& \int \di \sigma\, \sqrt{F^2(\r)\, \phi^{\prime\,2}+G^2(\r)\,\rho^{\prime\,2}}\,,
\eeqs
which describes a curve in the space $\left(\r(\sigma),\,\phi(\sigma)\right)$, with $0\leq \phi <2\pi$ and $\r \geq 0$, and
parameterised by the coordinate $\sigma$.
We assume that the functions $F$ and $G$ are both positive definite
and monotonically non-decreasing, which is true
in all the following applications. In this way
 large values of $\r$  correspond to 
short-distance physics in the dual quantum system.
We look for solutions that extend 
to a UV boundary $\r_U$ in the radial direction. The curve reaches the boundary at two points with angular separation $\bar{\phi}$,
and in the limit $\r_U\rightarrow +\infty$ it is orthogonal to the boundary.~\footnote{The 
presence of a boundary in the space means that we should also add a boundary-localised term to the action.
We do not write it explicitly, but implicitly we use it to remove
a divergence in the energy of the configurations.}

There exist several classes of solutions to the equations of motion.
The first type is given by $\phi^{\prime}=0$ and we refer to it as {\it disconnected}.
It consist of two straight lines extending from $\r_U$ all the way down into 
$\r\rightarrow 0$ at a constant 
angular separation.
By replacing into the action, their energy is
\beqs
E_0&=&2 \int_{0}^{\r_U} \di \r \,  G(\r)\,.
\eeqs

More interesting solutions have a U-shaped profile,
characterised by the value $\hat{\r}_o>0$ of the turning point:
 the solution at $\r=\hat{\r}_o$ obeys $\r^{\prime}(\sigma)/\phi^{\prime}(\sigma)=0$. 
We  refer to these solutions as {\it connected}. 

After deriving the equations of motion, we exploit reparametrization invariance
of the action to choose the ansatz $\r=\sigma$ and hence $\phi=\phi(\r)$. 
The connected configurations are described by two symmetric branches
joined smoothly at $\hat{\r}_o$.  
We define the following functions
~\cite{NPR,FPS}:
\beqs\label{Veff}
V^2_{eff}(\r,\hat{\r}_o) &\equiv& \frac{F^2(\r)}{G^2(\r)}\left(\frac{F^2(\r)}{F^2(\hat{\r}_o)}-1\right)\,,\\[2mm]
{\cal Z}(\r)&\equiv&\partial_{\r}\left(\frac{G(\r)}{\partial_{\r}F(\r)}\right)\label{Z}\,,
\eeqs
and the solution reads
\beqs
\phi(\r,\hat{\r}_o)&=&\left\{\begin{array}{cc}
\frac{\bar{\phi}}{2}\,-\,\int_{\hat{\r}_o}^{\r} \di \omega \,\frac{1}{V_{eff}(\omega,\hat{\r}_o)}\,,
&\left(x<\frac{\bar{\phi}}{2}\right)\cr
&\cr
\frac{\bar{\phi}}{2}\,+\,\int_{\hat{\r}_o}^{\r} \di \omega \,\frac{1}{V_{eff}(\omega,\hat{\r}_o)}\,,
&\left(x>\frac{\bar{\phi}}{2}\right)\cr
\end{array}\right.
\eeqs
The angular separation at the boundary and the energy of the configuration are given by
\beqs
\bar{\phi}(\hat{\r}_o)&=&2\int_{\hat{\r}_o}^{\r_U} \di \r\, \frac{1}{V_{eff}(\r,\hat{\r}_o)}\nonumber\\
&=&2\int_{\hat{\r}_o}^{\r_U} \di \r\, \frac{G(\r)}{F(\r)}\frac{1}{\sqrt{\frac{F^2(\r)}{F^2(\hat{\r}_o)}-1}}\,,\\[1mm]
E(\hat{\r}_o)&=&2\int_{\hat{\r}_o}^{\r_U} \di \r\,\frac{\di E}{\di \r}\nonumber\\
&=&2\int_{\hat{\r}_o}^{\r_U} \di \r\, \frac{F(\r)G(\r)}{F(\hat{\r}_o)}
\frac{1}{\sqrt{\frac{F^2(\r)}{F^2(\hat{\r}_o)}-1}}\,,
\eeqs
where the last expression has been obtained by replacing the classical solution into the action.

We remind the reader of three important facts about the U-shaped embeddings.
\begin{itemize}
\item The function $F(\r)$ must be positive definite and monotonically increasing for any $\r>0$, 
as visible from the definition of $V_{eff}$ and how it enters the solution to the equations of motion.
As we said, this will always be the case in this paper.

\item We must find that  $\lim_{\r\rightarrow +\infty} V_{eff}=+\infty$,
in order for the relevant boundary conditions to be satisfied.
Since we choose a parametrisation of the space in which
$\phi$ is compact, we  need a stronger condition:
the integral giving the asymptotic separation $\bar{\phi}(\hat{\r}_o)$ must converge.

\item The requirement ${\cal Z}<0$ yields a sufficient condition ensuring 
stability (the absence of tachyonic 
fluctuations of the classical configuration), and descends from the concavity conditions of
the relevant thermodynamic potential. 
By contrast, if ${\cal Z}>0$ for every $\r\geq 0$, then in the limit $\r_U\rightarrow +\infty$ 
all the U-shaped configurations are classically unstable.

\end{itemize}

Recall the general result
\beqs
\frac{\di E}{\di \bar{\phi}} &=& F(\hat{\r}_o)\,,
\eeqs
that does not depend either on whether one takes the limit $\r_U\rightarrow +\infty$ or on the fact that $E$ diverges in this limit and we must add a counterterm.  Such a counterterm cannot be dependent on $\hat{\r}_o$, 
and hence not on $\bar{\phi}$. 
This result indicates that
$F(\hat{\r}_o)$ is  the effective tension of the one-dimensional object described by $\cal S$.

We note two important things. 
First of all, the energy of the disconnected configuration 
does not depend on $\bar{\phi}$:
 there is a whole one-parameter class of
inequivalent disconnected configurations all of the same energy.
Secondly,  if  
 $\lim_{\hat{\r}_o\rightarrow 0}F(\hat{\r}_o)=0$, then the connected configuration with $\hat{\r}_o=0$ 
is indistinguishable from a disconnected configuration with the same value of $\bar{\phi}$.
This allows us to compare the energies of solutions in the two classes,
and  we can regulate and renormalise them in the same way.

The fact that $F$ is positive-definite, implies that the functions $E(\hat{\r}_o)$ and
$\bar{\phi}(\hat{\r}_o)$ are both either increasing or decreasing functions.
Moreover, since $F$ is monotonically increasing, given two 
solutions with different values of $\hat{\r}_o$ but the same value of $\bar{\phi}$,
necessarily $\di E/\di \bar{\phi}$ is larger for the solution with larger $\hat{\r}_o$.

Now, assume that there exist two different branches of connected solutions which, by appropriately varying
$\hat{\r}_o$, approach the point $(\bar{\phi}_1,E_1)$ from the same side, in the space of globally defined quantities.
Then, there is a neighbourhood of  $(\bar{\phi}_1,E_1)$  where the two branches 
represent inequivalent solutions for the same values of $\bar{\phi}$.
As we argued, the solutions on the branch 
which is controlled by larger values of $\hat{\r}_o$ will have larger values of  
$\di E/\di \bar{\phi}$.
Hence, they  have lower energy 
than the configurations with smaller $\hat{\r}_o$
 for $\bar{\phi}<\bar{\phi}_1$, and are thermodynamically favoured,
while they have higher energy  for $\bar{\phi}>\bar{\phi}_1$, in which case the  configurations 
having smaller $\hat{\r}_o$ are dominant.
This consideration will be useful in order to comment on the numerical results of our analysis.

We are now ready to tackle the main point we want to highlight. 
The parameter $\r_{U}$ is to be understood 
as a UV-cutoff and we shall choose it to be always much larger than any
dynamical scale  in the background.  We should take the limit $\r_U\rightarrow +\infty$
to recover physical results. 
Let us make explicit the cutoff dependence of the globally defined quantities $\bar{\phi}(\hat{\r}_o,\r_U)$
and $E(\hat{\r}_o,\r_U)$.
For large values of $\hat{\r}_o$ there are two possible orders of limits:
\begin{itemize}
\item[a)] we could first fix $\hat{\r}_o$, take the limit $\r_U\rightarrow +\infty$, and thereafter vary
 $\hat{\r}_o$ 
\beqs
\bar{\phi}_a&\equiv&\lim_{\hat{\r}_o\rightarrow +\infty} \lim_{\r_U\rightarrow +\infty} \bar{\phi}(\hat{\r}_o,\r_U)\,,\\
\bar{E}_a&\equiv&\lim_{\hat{\r}_o\rightarrow +\infty} \lim_{\r_U\rightarrow +\infty} 
\left(\frac{}{}E(\hat{\r}_o,\r_U)\,-\,E_0(\r_U)\right)\,,
\eeqs
\item[b)] alternatively, we could first keep fixed the UV cutoff, and study the system for varying $\hat{\r}_o$,
and only afterwards take the limit $\r_U\rightarrow \infty$
\beqs
\bar{\phi}_b&\equiv&\lim_{\r_U\rightarrow +\infty}  \lim_{\hat{\r}_o\rightarrow +\r_U}  \bar{\phi}(\hat{\r}_o,\r_U)\,=\,0\,,
\\
\bar{E}_b&\equiv&\lim_{\r_U\rightarrow +\infty}  \lim_{\hat{\r}_o\rightarrow +\r_U} 
 \left(\frac{}{}E(\hat{\r}_o,\r_U)\,-\,E_0(\r_U)\right)\,.
\eeqs
\end{itemize}

In most cases, the two limits commute, since frequently $\bar{\phi}_a=0=\bar{\phi}_b$.
Notice that to calculate $\bar{E}$ from $E$, we subtracted the divergent $E_0(\r_U)$, and in this scheme the disconnected solutions 
always have vanishing energy. 

The potential problem arises when $\bar{\phi}_a$ is finite.
There exist many examples of this, some of which are well known (for instance the 
D3-D7 system~\cite{KK}). Convergence of the limits defining $\bar{\phi}_a$ means
that, at least  for $\hat{\r}_o$ large enough, $\lim_{\r_U\rightarrow +\infty} \bar{\phi}(\hat{\r}_o,\r_U)$
becomes effectively independent of $\hat{\r}_o$.
By contrast, if we compute $\bar{\phi}_b$, 
for $\hat{\r}_o$ large enough the angular separation will first tend to converge towards $\bar{\phi}_a$ while increasing $\hat{\r}_o$, 
but when the configuration becomes
short and $\hat{\r}_0$ is taken close to $\r_U$, suddenly the angular separation starts to decrease and eventually vanish.
The resulting {\it short} configurations keep existing for any finite choice of $\r_U$, 
but they probe only a very narrow region close to the boundary. 
For this reason, these configurations cannot be seen 
if we take the ordering as in case a) above.

The question is then: are these short connected solutions, that are always localised in the proximity of the 
regulating $\r_U$, physical configurations, or a mere artefact of the regulation procedure?
Equivalently, which of the cases a) and b) yields physically sensible results?
The answer is that the short configurations are necessary in order to cure an otherwise
fatal pathology in the free energy, which may become discontinuous as a function of 
the control parameter $\bar{\phi}$ if we follow procedure a).
Hence we must use procedure b). We will see explicit examples of this discontinuity
in Section~\ref{Sec:chiralbreak}.

Short configurations are of no 
interest per se,  because they do not 
probe the geometry of the space described by the functions $F$ and $G$,
and hence are unrelated to the dual field theory.
Unfortunately,
they become thermodynamically favoured when $\bar{\phi}<\bar{\phi}_a$, inducing a phase transition.
For all practical purposes this amounts to the existence of a physical lower bound on $\bar{\phi}$,
below which the dynamics is dominated by cutoff effects.
This phenomenon is reminiscent of {\it bulk} phase transitions on the lattice,
and we adopt the convention of calling it by the same name.

Let us now explain why procedure b) yields a phase-transition when $\bar{\phi}_a>0$.
We first fix a large value of $\r_U$.
When varying $\hat{\r}_o$ in the region where the angular separation is approaching
$\bar{\phi}_a$, there are two possibilities. If $\di \bar{\phi}/\di \hat{\r}_o<0$, nothing special happens: $\bar{\phi}_b$
will decrease and approach $\bar{\phi}_a$, until $\hat{\r}_o$ becomes so close to
$\r_U$ that the configuration is short, and $\bar{\phi}_b$ keeps decreasing monotonically.
On the other hand, if $\di \bar{\phi}/\di \hat{\r}_o>0$ for some large $\hat{\r}_o$,
by increasing $\hat{\r}_o$ the sign of this derivative will have to change,
since ultimately $\bar{\phi}_b$ vanishes.
This signals a turning point in the $(\bar{\phi}_b,E_b)$ plane,
giving rise to two different branches of solutions.
As we said, in the proximity  of the turning point 
the derivative $\di E / \di \bar{\phi}$ will be larger for the branch with larger $\hat{\r}_0$.
This is the branch of the short configurations, which is hence thermodynamically favoured.

By looking at $E_b$ one sees that 
the short solutions 
have negative energy, which would diverge for $\r_U\rightarrow +\infty$: they dominate the dynamics 
atop any other configuration.
However, this branch does not exist for $\bar{\phi}>\bar{\phi}_a$.
By contrast the disconnected solution always exists and has vanishing energy, 
which means that 
at $\bar{\phi}=\bar{\phi}_a$ there must be a  phase transition,
with either the disconnected or a connected solution becoming the minimum of the energy
in the physical region $\bar{\phi}>\bar{\phi}_a$.

Let us finally explain why procedure a) is potentially problematic.
We can think of this setup as the saddle-point approximation of a 
statistical mechanics system, and interpret $\bar{E}$ as the Gibbs free energy,
expressed in terms of the external control parameter $\bar{\phi}$.
In the way we computed it earlier, $\bar{E}(\bar{\phi})$ is in general multivalued,
and the approximation requires us to retain only its global minimum,
when several different configurations sharing the same $\bar{\phi}$ exist.
The subtle point is that the resulting $\bar{E}(\bar{\phi})$ must be continuous,
and hence (global) minimization must be carried out
before taking the $\r_U\rightarrow +\infty$ limit.

The drawback of procedure a) is that, while the results it yields for all other branches agree with procedure b), it completely suppresses the short configurations.
Since such solutions have the lowest energy
(when $\bar{\phi}_a>0$), removing them may result
in unphysical discontinuities of the Gibbs free energy. Consequently, 
procedure a) does not provide a good approximation of the free energy.

In computing any physical observables, including the free
energy --- but also the separation $\bar{\phi}$ itself ---  
one must first find the global minimum of the energy, only afterwards apply subtractions and finally take the $\r_U\rightarrow +\infty$ limit.
Ultimately, following procedure b) ensures that $\bar{E}(\bar{\phi})$ is always continuous,
provides a natural explanation for the bound $\bar{\phi}>\bar{\phi}_a$,
and does not affect in any way phenomena taking place in the physical region above this bound.

\section{The wrapped-D5 system.}
\label{Sec:wrappedD5}

We focus on a special, large class of Type IIB backgrounds
which we refer to as the {\it wrapped-D5} system.
We do not repeat  all the details about the general solutions,
which can be found elsewhere~\cite{CNP,HNP}. We follow closely the notation in
the recent paper~\cite{FPS} and only mention those element that are directly relevant for the following sections.

The system  is a truncation of Type IIB supergravity that contains only gravity, the dilaton $\Phi$ and the RR three-form $F_3$. 
The internal part of the metric is related to $T^{1,1}$, and we write it in terms of the following vielbein:
\beqs
e_1&=&-\sin \theta\, \di \phi\,,\,\,\,\,\,\,\,\nonumber
e_2\,=\, \di \theta\,,\,\,\,\,\,\,\,\\
e_3 &=& \cos \psi\, \sin\tilde{\theta}\,\di \tilde{\phi}-\sin \psi\, \di \tilde{\theta}\,,\\
e_4 &=&\sin \psi \,\sin\tilde{\theta}\,\di \tilde{\phi}+\cos \psi \,\di \tilde{\theta}\,,\,\,\,\,\,\,\,\nonumber\\
e_5 &=&\di \psi +\cos{\theta}\,\di{\phi}+\cos\tilde{\theta}\,\di\tilde{\phi} \,,\nonumber
\eeqs
where the range of the five angles is
$0\leq \theta\,,\,\tilde{\theta}<\pi\,,\,
0\leq\phi\,,\,\tilde{\phi}<2\pi\,,\,0\leq\psi<4\pi$. We assume that the functions defining
the background depend only on the radial coordinate $\r$. 
The  metric (in Einstein frame) reads
\bea
\di s^2_E &=& \alpha' g_s \,e^{\Phi/2} \Big[ (\alpha' g_s)^{-1} dx_{1,3}^2 + ds_6^2 \Big], \nonumber\\[3mm]
\di s_6^2 &=& e^{2k}d\rho^2
+ e^{2 h}
(e_1^2 + e_2^2) \label{nonabmetric424}
\\
\nonumber
&+&\frac{e^{2 {g}}}{4}
\left(\frac{}{}(e_3+a\,e_1)^2
+ (e_4+a\,e_2)^2\right)
+ \frac{e^{2 k}}{4}e_5^2\,.
\eea

The string-frame metric is  $\di s^2 = e^{\frac{\Phi}{2}}\di s^2_E$,
we fix the coupling to $\alpha^{\prime}g_s=1$,
and ignore the non-vanishing $F_3$ (see for instance~\cite{HNP}). 

The system of BPS equations derived using this ansatz can be reorganized
 in terms of convenient  functions~\cite{HNP}:
\beqs
4 \,e^{2h}&=&\frac{P^2-Q^2}{P\cosh\tau -Q}\,, \qquad e^{2{g}}= P\cosh\tau -Q\,, \nonumber\\
e^{2k}&=& 4 \,Y\,,\qquad\qquad  \qquad a=\frac{P\sinh\tau}{P\cosh\tau -Q}\,,\;\;
\label{functions}
\eeqs
such that it reduces to a  single decoupled second-order equation for the function $P(\rho)$ that takes the form
\beq
P'' + P'\,\Big(\frac{P'+Q'}{P-Q} +\frac{P'-Q'}{P+Q} - 4 
\coth(2\rho-2{\rho}_0)
\Big)=0\,,
\label{Eq:master}
\eeq
where the prime refers to derivatives with respect to $\r$.
All other functions can be read from
\beqs
\nonumber
Q&=&(Q_0+ N_c)\cosh\tau + N_c \,(2\rho \cosh\tau -1)\,,\\
Y&=&\frac{P'}{8}\,,\qquad\qquad
 \nonumber
\cosh\t\,=\,\coth(2\r-2{\r_0})\,,\\
e^{4\Phi}&=&\frac{e^{4\Phi_0} \cosh(2{\rho_0})^2}{(P^2-Q^2) Y\sinh^2\tau}\,.
\label{BPSeqs}
\eeqs
We refer to Eq.~(\ref{Eq:master}) as the {\it master equation}.
In the following we take the end of space to be at $\r_0=0$ 
and tune $Q_0=-N_c$. The remaining constant
 $\Phi_0$ is a free parameter of little physical meaning, and we  set  $\Phi_0=0$.
The generic solution for $P$  depends on two integration constants.

The master equation has to be solved numerically, the only solution known 
in closed form being the famous CVMN solution~\cite{MN,CV}:
\beqs
\hat{P}&=&2N_c \r\,.
\eeqs
All other solutions with regular $P$ 
are approximated by 
\beqs
P_a&\simeq&\sup\left\{\frac{}{} c_0\,,\,2N_c \r\,,\,3c_+ e^{4\r/3}\right\}\,,
\eeqs
where $c_0$ and $c_+$ are two integration constants.
When $c_0$ effectively vanishes, for small $\r$
the solution is best approximated  by the expansion  \cite{HNP}:
\beqs
P_{\ell}&=&h_1\r +\frac{4h_1}{15}\left(1-\frac{4N_c^2}{h_1^2}\right)\r^3 + {\cal O}(\r^5)\,,
\eeqs
in which $h_1\geq 2 N_c$ is the remaining integration constant. 
 The value $h_1=2N_c$ reproduces exactly the CVMN solution.
Discussions of  these backgrounds
 can be found elsewhere~\cite{FPS,EGNP,HNP}.
 In particular, in~\cite{EGNP} one finds the details about the
 relation between these backgrounds and the baryonic branch~\cite{BGMPZ}
 of the Klebanov--Strassler system~\cite{KS} and other backgrounds
 within the Papadopoulos--Tseytlin ansatz~\cite{PT} and its generalisation~\cite{consistentconifold},
 via the rotation (or U-duality) procedure introduced and studied in~\cite{MM,CNPZ,GMNP}.

\subsection{Flavored Abelian solutions.}

We introduce a related class of Type IIB solutions, which we refer to as Abelian~\cite{HNP}.
Some of the background functions 
(related to the gaugino condensate) are omitted, in particular the function $a$, which is related to the 
$SU(2)$-twisting from which the distinction between Abelian and non-Abelian backgrounds originates~\cite{CV}.

The Abelian solutions can be obtained by replacing $\tau=0$ in the earlier ansatz.
The only delicate piece is the dilaton, that in our conventions reads
\beqs
e^{4\Phi}&=&\frac{e^{4\Phi_0} e^{4\r}}{4(P^2-Q^2) Y}\,.
\label{BPSeqsA}
\eeqs
A further modification of the system consists in introducing flavor in the form of $N_f$ smeared D5-branes.
The details 
can be found in~\cite{CNP,HNP}.
The background metric obeys the same 
ansatz. 
The only modifications appear in the functions $Q$ and $Y$ and in the master equation:
\begin{widetext}
\beqs
Y&=&\frac{P'+N_f}{8}\,,\qquad\qquad 
Q\,=\,Q_o+N_c-\frac{N_f}{2}+\left(N_c-\frac{N_f}{2}\right)(2\r-1)\,,\\
0&=&P'' + (P'+N_f)\,\Big(\frac{P'+Q'+2N_f}{P-Q} +\frac{P'-Q'+2N_f}{P+Q} - 4 
\Big)\,.
\eeqs
\end{widetext}
For $N_f=2N_c$, 
there exists the exact solution~\cite{HNP}
\beqs\label{fullsol}
Q&=&\frac{3N_c}{4}\,,\qquad\qquad
P\,=\,\frac{9N_c}{4} +c_+ \,e^{4\r/3}\,,
\eeqs
with $c_+\ge0$ a constant. The end-of-space in the IR is now located at $\r\to-\infty$. Indeed, for $c_+=0$
all the background functions become constant, with the exception of the linear dilaton.
With abuse of language, we call it
{\it scale invariant}. 

\section{D7 embedding.}
\label{Sec:D7embedds}

We  embed a probe-D7~\cite{A,KuSo,DKS} so that it fills 
the four Minkowski coordinates  
and the internal three-manifold 
$(\tilde{\theta}, \tilde{\phi}, \psi)$. 
Since $B_2$ is trivial, we set the gauge field on the brane to ${\cal F}_2=0$. We treat $\rho$, $\phi$ and $\theta$ as functions of the embedding coordinate $\sigma$.
Integrating the rest of the angular variables, the DBI action 
${\cal S}_{D7} \propto \int \di^8 x\,\sqrt{-\tilde{g}_8}\, e^{-\Phi}$ 
becomes
\begin{widetext}
\beqs
{\cal S}_{D7} &\propto&\int \di^4 x\,  \di \sigma \, \sqrt{e^{4g+4k+6\Phi}\r^{\prime\,2}+e^{4g+2k+2h+6\Phi}\left(\theta^{\prime\,2}+\sin^2\theta\,\phi^{\prime\,2}\right)}\,.
\eeqs
\end{widetext}
The function $a$ disappears from this expression,  
so that we can use the same
formalism to deal both with Abelian and non-Abelian backgrounds.
For symmetry reasons, without loss of generality 
we  choose configurations at the equator $\theta=\frac{\pi}{2}$.
Finally, 
the action takes the form of ${\cal S}$ presented in Eq.~(\ref{genaction}), provided we make the identifications
\begin{equation}
F^2\,=\,e^{4g+2k+2h+6\Phi}\,,\qquad\qquad G^2\,=\,e^{4g+4k+6\Phi}\,.
\end{equation}

\subsection{Flavored Abelian backgrounds.}

We consider the flavoured Abelian solution, which is simple enough that we can perform most of the computations analytically, serving for illustrational purposes. The background functions read:
\beqs
F^2&=&\frac{e^{6\r}}{2}\,\sqrt{\frac{P-Q}{2(P^{\prime}+N_f)(P+Q)}}\nonumber\\
&=&\frac{\sqrt{3} \,e^{6\r}}{4\sqrt{6+2c_+\,e^{4\r/3}}}\,,\\
G^2&=&\frac{e^{6\r}}{(P+Q)^2}\,\sqrt{\frac{(P^{\prime}+N_f)(P^2-Q^2)}{2}}\nonumber\\
&=&\frac{e^{6\r}}{\sqrt{6}}\,\frac{3+2c_+\,e^{4\r/3}}{(3+c_+\,e^{4\r/3})^{3/2}}\,,
\eeqs
where we replaced the special solutions presented earlier, and fixed $N_c=1$.
Both $F$ and $G$ diverge at large-$\rho$,  as anticipated.

We start from the case $c_+=0$, for which 
\beqs
F^2&=&\frac{e^{6\r}}{4\sqrt{2}}\,,\qquad\qquad
G^2\,=\,\frac{e^{6\r}}{3\sqrt{2}}\,.
\eeqs
As a result, we find  
\beqs
V_{eff}^2&=&\frac{3}{4}\left(\frac{}{}-1+e^{6(\r-\hat{\r}_0)}\right)\,,\\[2mm]
\frac{\di E}{\di \r}&=&\frac{1}{2^{1/4}\sqrt{3}}\,\frac{e^{6\r}}{\sqrt{e^{6\r}-e^{6\hat{\r}_o}}}\,,\\[2mm]
{\cal Z}&=&0\,.
\eeqs
We immediately see that the conditions we outlined for $F$ and $V_{eff}$ are satisfied.
The condition for ${\cal Z}$ is more subtle: its vanishing is a limiting case.

\begin{figure}
\begin{center}
\begin{picture}(180,110)
\put(0,0){\includegraphics[height=3.2cm]{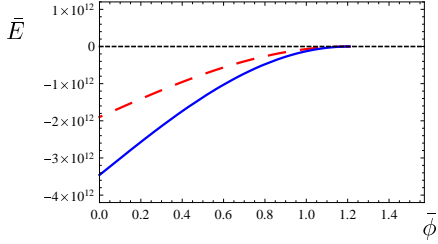}}
\end{picture} 
\caption{Dependence of the subtracted energy $\bar{E}$ on the angular separation $\bar{\phi}$,
computed for the scale-invariant solutions of the Abelian flavoured system.
We show the disconnected configuration (black short-dashed), together with the connected configurations 
computed with fixed $\r_U$ by varying $\hat{\r}_o$. The (red) long-dashed
 curve has $\r_U=9.8$, while the (blue) continuous curve 
was computed with $\r_U=10$.}
\label{Fig:plotabelianlinear}
\end{center}
\end{figure}

Integrating one obtains the energy and asymptotic separation
\beqs
\bar{\phi}(\hat{\r}_o,\r_U)&=&\frac{4}{3\sqrt{3}}\arctan \sqrt{-1+e^{6(\r_U-\hat{\r}_o)}}\,,\\
E(\hat{\r}_o,\r_U)&=&\frac{2^{3/4}}{3\sqrt{3}}\sqrt{e^{6\r_U}-e^{6\hat{\r}_o}}\,,\\
E_0&=&\frac{2^{3/4}}{3\sqrt{3}}\frac{}{}e^{3\r_U}\,,
\eeqs
where the lower end of the integral yielding $E_0$ is the end-of-space $\r\rightarrow -\infty$.

\begin{figure}[h]
\begin{center}
\begin{picture}(180,320)
\put(0,0){\includegraphics[height=9.8cm]{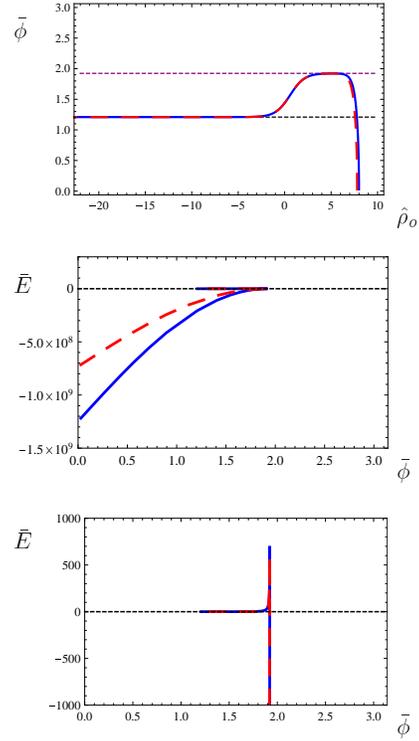}}
\end{picture} 
\caption{Numerical study in the Abelian flavoured case with $c_+=1=N_c$, and $N_f=2$.
The (blue) continuous curves show the results  for the connected configurations with $\r_U=8$,
and the (red) long-dashed curves for $\r_U=7.8$.
The top panel shows the dependence of $\bar{\phi}$ on $\hat{\r}_0$ in the two cases, the short-dashed lines
represent the asymptotic values $\bar{\phi}=2\pi/3\sqrt{3}$ (black) and $\bar{\phi}=\sqrt{6}\pi/4$ (purple).
The middle plot shows the dependence of the subtracted energy $\bar{E}$ on $\bar{\phi}$, with the 
short-dashed line (black) representing the disconnected solution. The bottom panel shows a detail of the middle one.}
\label{Fig:plotabelianexp}
\end{center}
\end{figure}

Following procedure a), i.~e. if we first take the limit in which we remove the cutoff,
we see the emergence of a peculiar situation.
Namely, independently of $\hat{\r}_o$, we find that 
\beqs
\lim_{\r_U\rightarrow +\infty}\bar{\phi}(\hat{\r}_o,\r_U)&=&\frac{2\pi}{3\sqrt{3}}\,\equiv\,\bar{\phi}_a\,,\\
E(\hat{\r}_o,\r_U)-E_0(\r_U)&=&-\frac{e^{6\hat{\r}_o-3\r_U}}{2^{1/4}3\sqrt{3}}+\cdots
\rightarrow 
 0\,.
\eeqs
There exists only one point in the $(\bar{\phi}_a,\bar{E}_a)$ plane.
All the connected configurations admissible as solutions, for any $\hat{\r}_o$, yield the same 
result in terms of physical quantities.
To this, one has to add the disconnected configurations: given that $\lim_{{\r}\rightarrow -\infty}F(\r)=0$,
the limit $\hat{\r}_o\rightarrow -\infty$ is 
indistinguishable from the disconnected configuration.

For any choice of $\bar{\phi}$, there exists the disconnected
configuration, that has constant energy. For the special value $\bar{\phi}=\bar{\phi}_a$ there exists a whole other branch of connected solutions, again with the same energy.

However, following procedure b) we obtain another branch of configurations, the short connected ones.
We display in Fig.~\ref{Fig:plotabelianlinear}  
the behaviour of the subtracted energy $\bar{E}=E-E_0$, as a function of $\bar{\phi}$,
for two different values of the cutoff $\r_U=9.8$ and $\r_U=10$,
compared with the disconnected configuration.~\footnote{Here and in the following, the plots have been made by fixing a large value of the UV cutoff, and then subtracting the energy of the disconnected solution for the same $\r_U$.  The choice of cutoff is large enough, that changing it to even larger values, has no appreciable effect on the side of the phase transition connected to the gravity dual.  On the other side of the bulk transition, by looking at increasing values of the cutoff, one sees that the energy eventually diverges to $-\infty$.  The plots of the different values of $\r_U$ (see Fig.~\ref{Fig:plotabelianlinear} and Fig.~\ref{Fig:plotabelianexp}) are used to illustrate this.}

Both from the analytical results, and from the figure, we can see four important things.
First, the short configurations have lower energy than the disconnected ones,
and hence are always dominant for $\bar{\phi}<\bar{\phi}_a$.
Secondly, rather than decoupling when $\r_U\rightarrow +\infty$, the energy of the short configurations
keeps decreasing, so that they become more dominant
the larger the value of $\r_U$.
Third, the limit $\bar{\phi}\rightarrow\bar{\phi}_a$ coincides with the taking the limit $\hat{\rho}_0\rightarrow-\infty$.
Since in this limit $F(\hat{\r}_o)\rightarrow 0$,  
 the branch of short configurations joins smoothly with the disconnected one,
giving rise to a second-order phase transition.
Finally,  the short connected solutions are concave functions of $\bar{\phi}$,
which means that they are classically stable.

The algebra is simple enough that we can 
solve  for $\hat{\r}_o$ in $\bar{\phi}(\hat{\r}_o,\r_U)$, and replace into the subtracted energy:
\beqs
\bar{E}(\bar{\phi},\r_U)&=&\frac{2^{3/4}}{3\sqrt{3}}\,e^{3\r_U}\,\left(\sin\left(\frac{3\sqrt{3}}{4}\bar{\phi}\right)-1\right)\,,\\
\partial_{\bar{\phi}}\bar{E}(\bar{\phi},\r_U)&=&\frac{1}{2^{5/4}}\,e^{3\r_U}\,\cos\left(\frac{3\sqrt{3}}{4}\bar{\phi}\right)\,,\\
\partial^2_{\bar{\phi}}\bar{E}(\bar{\phi},\r_U)&=&-\frac{3\sqrt{3}}{2^{1/4} 8}\,e^{3\r_U}\,\sin\left(\frac{3\sqrt{3}}{4}\bar{\phi}\right)\,.
\eeqs 
In the limit $\bar{\phi}\rightarrow \bar{\phi}_a$,
we find $\bar{E}=0=\partial_{\bar{\phi}}\bar{E}$: the free  energy is continuous and has 
continuous derivative, but $\partial^2_{\bar{\phi}}\bar{E}\propto e^{3\r_U}\neq 0$, so the system exhibits a second-order phase transition.

We now consider the full solution (\ref{fullsol}) and set $c_+=1$ for convenience. In the IR the geometry  
reproduces the previous results. In the UV new effects appear. 
Fig.~\ref{Fig:plotabelianexp} shows the numerical study of the system, 
for which we used the cutoffs $\r_U=7.8$
and $\r_U=8$. 

Plugging (\ref{fullsol}) into (\ref{Veff}) and (\ref{Z}), it can be seen that $V_{eff}$ diverges when $\r\rightarrow +\infty$.
However, we encounter a problem with the function ${\cal Z}$, which is positive definite.
We expect the concavity condition to be violated, and hence
the long connected configurations are classically unstable.
This is  visible in the bottom panel of Fig.~\ref{Fig:plotabelianexp}, where the branch of connected, long  solutions, which always have positive energy,
 has the wrong convexity.

If we were to restrict our attention to connected configurations, and apply procedure a),
thus ignoring the short configurations, we would solely find the branch with positive $\bar{E}$. This would pose two puzzling results.
The first is that these solutions are classically unstable.
The second is that these configurations only exist provided 
\beqs
\frac{2\pi}{3\sqrt{3}}<\bar{\phi}<\frac{\sqrt{6}}{4}\pi\,.
\eeqs
The asymptotic values are reached
for $\hat{\r}_o\rightarrow -\infty$ in the case of the lower bound,
and for $\hat{\r}_o$ asymptotically large, but still far from the UV cutoff, in the case of the upper bound.
The presence of the lower bound is not a surprise: we have the same geometry as before, in the deep IR,
and hence brane configurations that extend very deep in the IR effectively behave as in
the scale-invariant case. 

The upper bound comes from the drastic modification of the geometry  
induced in the far-UV by $c_+>0$. 
We can approximate the geometry here by setting $Q=0=N_f$
and $P=e^{4\r/3}$. By replacing in the various functions needed, we find
\beqs
\bar{\phi}(\hat{\r}_o,\r_U)&=&\sqrt{\frac{3}{2}}\arctan\sqrt{\frac{}{}-1+e^{\frac{16}{3}(\r_U-\hat{\r}_o)}}\,,
\eeqs
which for $\r_U\rightarrow +\infty$ converges to $\bar{\phi}(\hat{\r}_o,\r_U)\rightarrow \sqrt{6}\pi/4$.

On the basis of the general considerations we are making, 
we expect that procedure b) reveals another branch of solutions, which exists
for any value $\bar{\phi}<\bar{\phi}_a=\sqrt{6} \pi/4$.
This can be seen in the figure, which also shows that the short connected configurations become more dominant 
when the UV cutoff is taken to be large.

The transition between short and disconnected configurations is first order,
contrary to the scale-invariant case.
Another distinction is that long connected configurations do not have the same energy as the disconnected ones. 
Long connected solutions, besides being classically unstable, always
have energy larger than the short connected or the disconnected solutions.

The conclusion is that in this system there is now a bound $\bar{\phi}>\sqrt{6}\pi/4$
on the control parameter.  Furthermore, the disconnected configuration is always the 
one physically realised, so this type of set-up cannot be used to describe chiral symmetry breaking.

\section{Chiral symmetry breaking in the wrapped-D5 system.}
\label{Sec:chiralbreak}

We move now to the main topic of the paper, that is, the unflavored, non-Abelian solutions to the wrapped-D5 system,
for which $P_a$ or $P_{\ell}$ provide an approximation of the background.
There exist only two admissible UV-asymptotic expansions: either $P$ is linear with $\r$, or exponentially growing.
In the former case, $\bar{\phi}$ converges towards a vanishing value for connected configurations
when $\hat{\r}_o$ is large. 
All possible backgrounds of this class are not affected by bulk transitions  
and the existence of a phase transition between connected and disconnected
configurations has been studied in~\cite{FPS}.

We hence focus on solutions for which the asymptotic
behaviour is dominated by the exponential growth of $P$.
There are several subclasses, depending on the IR
behaviour of the background.

\subsection{Non-Abelian solutions and the baryonic branch.}

We start with the one-parameter family of regular backgrounds with $c_0=0$,
i.e. those that are controlled by a linear $P$ for $\r<\bar{\r}$ and exponential for $\r>\bar{\r}$.
It has been noted in~\cite{MM,CNPZ,GMNP} that after U-duality these reproduce the baryonic branch of the Klebanov--Strassler system.
With abuse of language we refer to them as the baryonic branch,
although strictly speaking they agree with it only in the IR. 
In the deep IR they are well approximated 
by the expansion  $P_{\ell}$.
The parameter $c_+$ in 
the UV expansion controls the scale 
$\bar{\r}$ below which the 
dimension-two  baryonic VEV becomes important.

\begin{figure}
\begin{center}
\begin{picture}(180,220)
\put(0,0){\includegraphics[height=6.6cm]{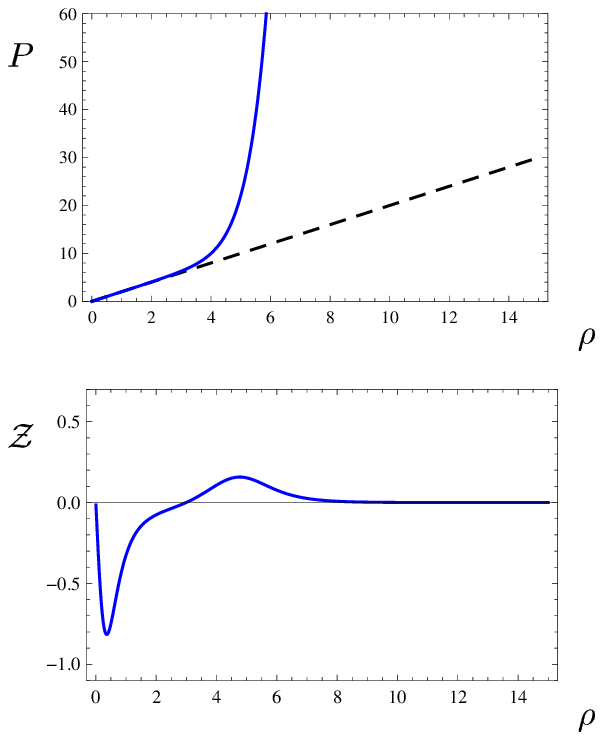}}
\end{picture} 
\caption{Example of a numerical solution to the non-Abelian system,
with IR asymptotics given by $P_{\ell}$.
The top panel shows the function $P$, the bottom panel the function ${\cal Z}$.
We set $N_c=1=e^{\Phi_o}$.}
\label{Fig:baryonicfunctions}
\end{center}
\end{figure}

\begin{figure}
\begin{center}
\begin{picture}(180,420)
\put(0,0){\includegraphics[height=13cm]{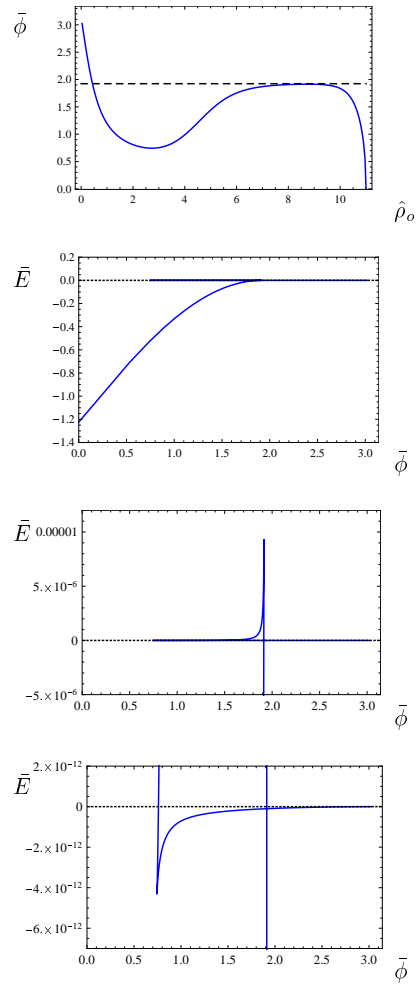}}
\end{picture} 
\caption{Study of the embedding for the background in Fig.~\ref{Fig:baryonicfunctions}.
The continuous (blue) curves are obtained for $\r_U=11$.
The top panel shows the functional dependence of $\bar{\phi}$ on $\hat{\r}_o$.
The dashed (black) line is the asymptotic value $\bar{\phi}=\sqrt{6}\pi/4$.
The other panels show the dependence of $\bar{E}$ on $\bar{\phi}$, with
 the dashed (black) line representing the disconnected solutions.}
\label{Fig:baryonicbranes}
\end{center}
\end{figure}

We display in Fig.~\ref{Fig:baryonicfunctions} an example in this class,
with $\bar{\r}\simeq 3$ (obtained with $h_1=2.004$).
The figure shows the functions $P$ and $\cal Z$.
The necessary conditions outlined in the general discussion on
the functions $F$ and $V_{eff}$ are satisfied, but the function ${\cal Z}$ is not negative definite.
In particular, for $\r>\bar{\r}$ it becomes positive, indicating that we expect the arising of 
an instability.
We show in Fig.~\ref{Fig:baryonicbranes} the results of the numerical study
of the background in Fig.~\ref{Fig:baryonicfunctions}.

In this case there is a physical end-of-space in the IR, which we set to $\r_{0}=0$.
In this region, the geometry resembles that of the CVMN solution, and indeed as $\hat{\r}_o\rightarrow 0$, 
 the separation tends to $\bar{\phi}\rightarrow \pi$, while the energy $\bar{E}\rightarrow 0$. Moreover, $F(0)=0$
implies that  $\di \bar{E}/ \di \bar{\phi}=0$.
 Hence, the connected configurations that extend deep into the IR
 are the equilibrium solutions, having the correct concavity and energy lower than the disconnected ones.
 This situation is common to the whole class with IR asymptotics given by
 $P_{\ell}$.

 Fig.~\ref{Fig:baryonicbranes} shows details of the $(\bar{\phi},\bar{E})$ plane,
showing the four branches of solutions, their concavity, and 
 the range of $\bar{\phi}$ for which they exist. The branch that reaches the maximal separation 
 $\bar{\phi}\rightarrow \pi$ is clearly visible in the last panel of the figure.
It exists for  $\bar{\phi}>\bar{\phi}_1$, determined by the value of $\bar{\r}$
 (in the figure, $\bar{\phi}_1\simeq 0.7$).
A second branch is visible in the third panel from the top.
It has the wrong concavity, and hence represents unstable classical solutions that probe the geometry
no further than $\bar{\r}$. It exists only in the range
$\bar{\phi}_1<\bar{\phi}<\bar{\phi}_a=\sqrt{6}\pi/4$. 
Solutions along this branch always have energy larger than the first class.
The third class of solutions are the disconnected ones.

If only these three classes existed, we would be faced with a serious problem.
Interpreting the minimum of $\bar{E}$ as the free energy $\mathcal{G}$, we would be left with a discontinuity in $\mathcal{G}$ as a function of $\bar{\phi}$.
This is the result one gets by applying procedure a) while regulating the calculations. This example clearly shows that it yields nonsensical results.
Procedure b) allows us to retain a fourth branch consisting of short configurations localized in the vicinity of the UV cutoff. How close can be seen in the right-hand side of the top panel in  
Fig.~\ref{Fig:baryonicbranes}.
This is the branch that reaches $\bar{\phi}\rightarrow 0$, 
apparent in the second panel from the top of the figure.
It has the correct concavity, large values of $\di \bar{E}/\di \bar{\phi}$
(actually divergent in the $\r_U\rightarrow +\infty$ limit) and exists only for $\bar{\phi}<\bar{\phi}_a$. Whenever present it is always the minimum of $\bar{E}$.

The conclusion is that a first-order bulk phase transition takes place 
at $\bar{\phi}=\bar{\phi}_a$.
For smaller values of the control parameter, the equilibrium configurations are the short ones,
that do not probe the interior of the geometry.
For larger values the equilibrium is reached by long connected configurations with $\hat{\r}_o<\bar{\r}$. 

An alternative phrasing is that 
there exists a physical bound on the control parameter $\bar{\phi}>\bar{\phi}_a$. For smaller values it is not possible to 
interpret the results in dual field-theoretic terms.

Notice that, for any choice of $\bar{\r}$, the turning point in the $(\bar{\phi},\bar{E})$
plot is always situated at $\bar{\phi}_1<\bar{\phi}_a$. This is of crucial importance,
since it proves that $\bar{E}$ is continuous.
The position of the turning point depends on the value of $\bar{\r}$,
or equivalently $h_1$. It moves towards  $\bar{\phi}_1\rightarrow 0$ in the limit $\bar{\r}\rightarrow+ \infty$
(corresponding to $h_1\rightarrow 2$), while $\bar{\phi}_1\rightarrow \bar{\phi}_a$ when $h_1\rightarrow +\infty$.
The latter is the limit in which the baryonic VEV is suppressed and the rotation allows us to reconstruct the 
Klebanov--Strassler solution~\cite{GMNP,EGNP}.

We checked numerically that even for $h_1\simeq 40000$  
the turning point is at $\bar{\phi}_1<\bar{\phi}_a$.
There is a simple reason for this: solutions with $\hat{\r}_o\gg \bar{\r}$ still have $\di \bar{E}/\di \bar{\phi}>0$,
which requires that the turning point verifies $\bar{\phi}_1<\bar{\phi}_a$.

\subsection{Walking solutions on the baryonic branch.}

We now consider a non-vanishing $P$ in the IR, that is $P\simeq c_0$ for small values of $\r$, while in the far UV it is again exponentially growing as $P\simeq 3\,c_+e^{4\r/3}$.
There are two integration constants, $c_0$ and $c_+$, which are chosen such that there is an intermediate range where $P$ approximates the CVMN solution.

\begin{figure}[h]
\begin{center}
\begin{picture}(180,210)
\put(12,5){\includegraphics[height=6.6cm]{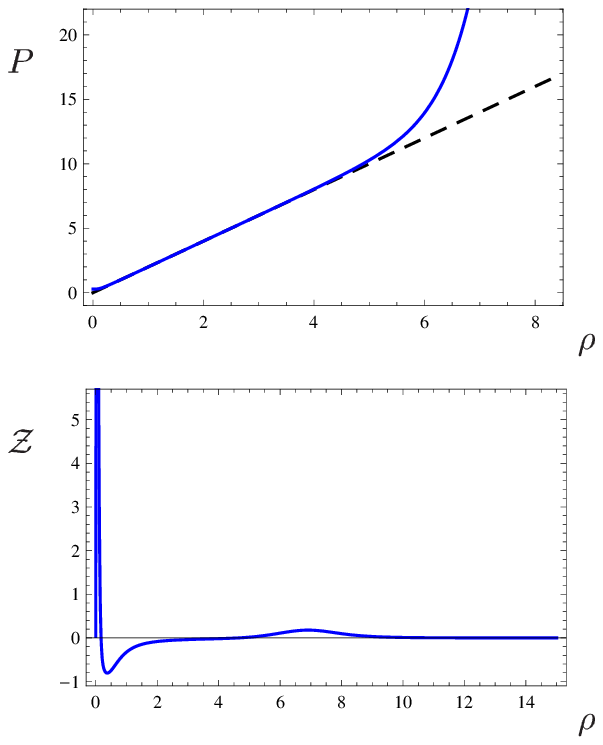}}
\end{picture} 
\caption{Example of a numerical solution to the non-Abelian system,
with IR asymptotics given by a constant $P\simeq c_0$.
The top panel shows the function $P$, the bottom panel the function ${\cal Z}$.
We set $N_c=1=e^{\Phi_o}$. Notice how the scale $\r_{\ast}$ is quite small.}
\label{Fig:baryonicfunctionsW}
\end{center}
\end{figure}

In Fig.~\ref{Fig:baryonicfunctionsW}, we show an example of $P$ in this class.
As can be seen, there are now three physical scales: the end of space $\r_0=0$,
a scale $\r_{\ast}$ below which $P$ is approximately constant (we call this the {\it walking} region),
and a scale $\bar{\r}$ above which $P$ is exponentially growing.
In the range $\r_{\ast}<\r<\bar{\r}$ we have that $P\simeq \hat{P}$.~\footnote{Unfortunately, there is neither a closed relation between $c_0$ and $c_+$ and the scales,
nor an easy way to extract their values from the numerics. Nevertheless, variation
in those parameters do not show qualitative changes in the features we uncover other
than the ones we comment.}

Now ${\cal Z}$ changes sign twice: deep in the IR and far in the UV,
${\cal Z}>0$, which suggests that probe branes extending only in the far UV, or very deep in the IR,
are classically unstable.
From the figure we identify $\bar{\r}\simeq 4$ and $\r_{\ast}\simeq 0.2$ with the zeros of ${\cal Z}$.

\begin{figure}[h]
\begin{center}
\begin{picture}(200,494)
\put(0,0){\includegraphics[height=16cm]{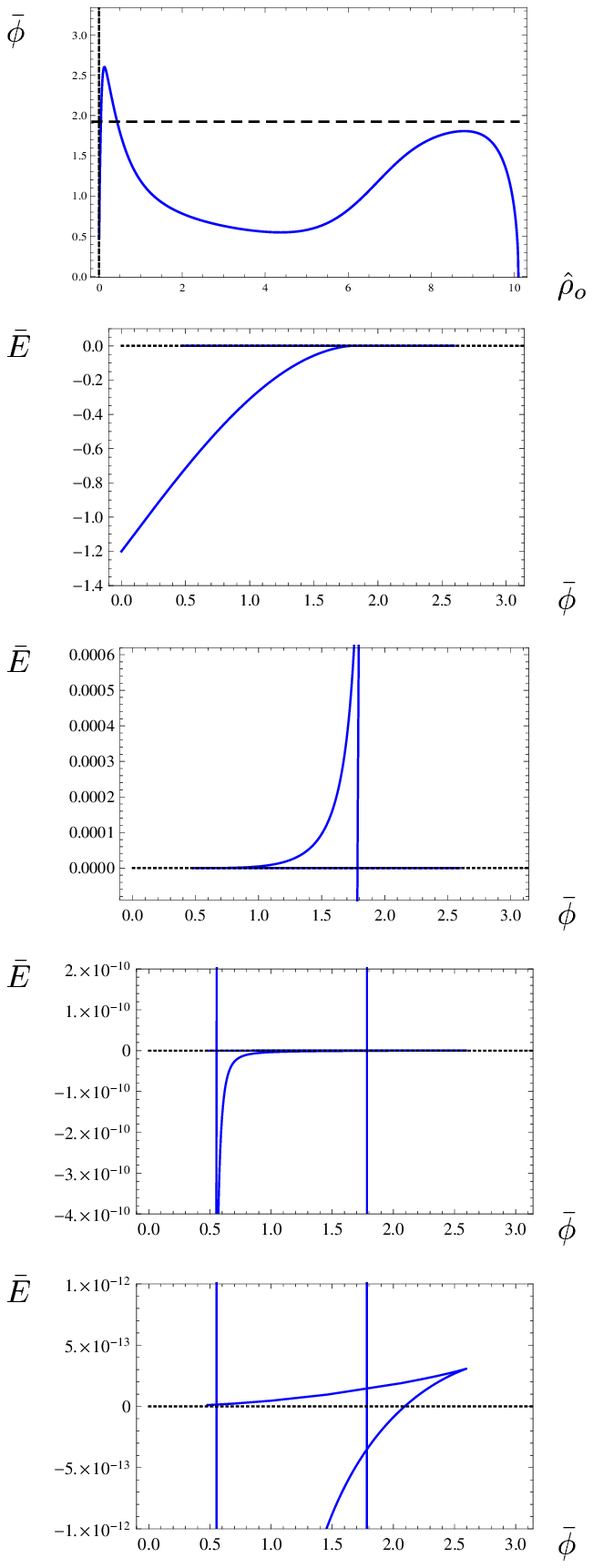}}
\end{picture} 
\caption{Study of the embedding for the background in Fig.~\ref{Fig:baryonicfunctionsW}.
The continuous (blue) curves are obtained for $\r_U=10.1$.
The top panel shows the functional dependence of $\bar{\phi}$ on $\hat{\r}_o$.
The long dashed (black) line is the asymptotic value $\bar{\phi}=\sqrt{6}\pi/4$ and 
the short dashed (black) line represents the
disconnected solutions.
The other panels show the dependence of $\bar{E}$ on $\bar{\phi}$.}
\label{Fig:baryonicbranesW}
\end{center}
\end{figure}

\begin{figure}[h]
\begin{center}
\begin{picture}(180,200)
\put(0,0){\includegraphics[height=6.5cm]{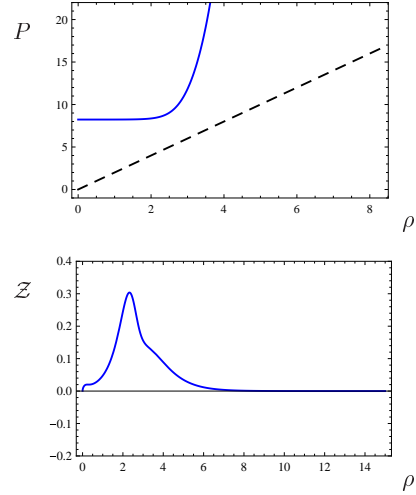}}
\end{picture} 
\caption{Example of a numerical solution to the non-Abelian system,
with IR asymptotics given by a constant $P\simeq c_0$.
The top panel shows the function $P$, the bottom panel the functions ${\cal Z}$.
We set $N_c=1=e^{\Phi_o}$.}
\label{Fig:functionsW}
\end{center}
\end{figure}

\begin{figure}[h]
\begin{center}
\begin{picture}(180,205)
\put(0,0){\includegraphics[height=6.4cm]{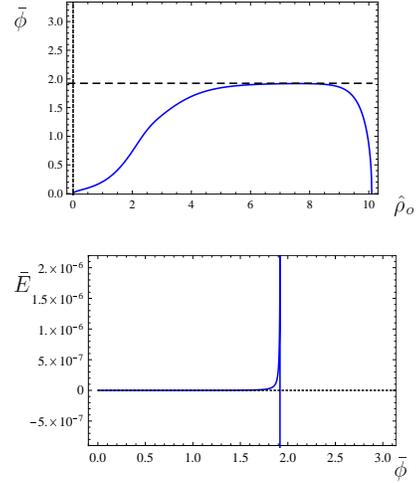}}
\end{picture} 
\caption{Study of the embedding for the background in Fig.~\ref{Fig:functionsW}.
The continuous (blue) curves are obtained for $\r_U=10.1$.
The top panel shows the functional dependence of $\bar{\phi}$ on $\hat{\r}_o$.
The long dashed (black) line is the asymptotic value $\bar{\phi}=\sqrt{6}\pi/4$ and
the short dashed (black) line represents the disconnected solutions.
The bottom panel shows the dependence of $\bar{E}$ on $\bar{\phi}$. 
}
\label{Fig:branesW}
\end{center}
\end{figure}

We embed D7-branes as explained in Section~\ref{Sec:D7embedds}.
The result is illustrated in Fig.~\ref{Fig:baryonicbranesW}, 
where we performed the calculations with UV cutoff $\r_U=10.1$.
First of all, for $\r>\r_{\ast}$ the background is essentially identical to the baryonic branch of the 
previous subsection. Unsurprisingly the probes behave in the same way as long as
$\hat{\r}_o>\r_{\ast}$. For branes that extend very close to the end of space,
with $\hat{\r}_o<\r_{\ast}$, the aforementioned classical instability manifests itself in the presence of a fifth branch,
along which the function $\bar{E}(\bar{\phi})$
has the wrong concavity. From the top panel of the figure one sees that there is a range of $\bar{\phi}$ where
we have five different branches of classical solutions (including the disconnected ones)  
yielding a richer structure of phase transitions.

The short connected configurations dominate the 
dynamics for $\bar{\phi}<\bar{\phi}_a$.
Increasing $\bar{\phi}$, there is a bulk phase transition, and in the range  $\bar{\phi}_a<\bar{\phi}<\bar{\phi}_c$
the connected configurations that 
we saw taking over in the baryonic branch minimize the free energy.
Nonetheless, this branch does not extend to $\bar{\phi}\rightarrow \pi$ any more: once $\hat{\r}_o<\r_{\ast}$,
the new, unstable branch appears, along which asymptotically $\bar{\phi}\rightarrow 0$ for $\hat{\r}_o\rightarrow 0$.~\footnote{In the bottom four panels of Fig.~\ref{Fig:baryonicbranesW}, the fifth (unstable) branch extends to meet the disconnected branch at $(\bar{E},\bar{\phi})=(0,0)$, but this feature is not shown due to limitations of the numerics.}
For large values $\bar{\phi}>\bar{\phi}_c$ it is now the disconnected solution that dominates the dynamics.
Hence there is a physically interesting phase transition between connected and disconnected solutions.

This result reproduces the behaviour expected for solutions with CVMN
asymptotics discussed in~\cite{FPS}, but only provided $\r_{\ast}$ is small.
By looking at the top panel of Fig.~\ref{Fig:baryonicbranesW}, we see that 
if $\r_{\ast}\gsim 0.5$, all the connected configurations yield $\bar{\phi}<\bar{\phi}_a$.
This is the regime dominated by the short solutions.
When $\r_{\ast}>0.5$, the disconnected configurations exist solely, with $\bar{\phi}>\bar{\phi}_a$.
The physics of the D7 probes is trivial in this case, so we show only the small $\r_{\ast}$ examples.

For completeness, we reconsider the case in which $P\gg \hat{P}$ for every $\r$.
This has already been discussed in~\cite{FPS}, where it is shown that connected configurations
are always unstable, and the disconnected ones have always lower energy.
Fig.~\ref{Fig:functionsW} displays an example in this class.
Notice that ${\cal Z}$ is positive definite.
Taking into account the short connected solutions
has the  effect of restricting the physical range of $\bar{\phi}$.
Yet, this does not affect the conclusion of~\cite{FPS}, according to which 
in these backgrounds only the disconnected configurations 
are physically realised, meaning that the D7 system
cannot describe chiral-symmetry breaking.
We show this fact explicitly in Fig.~\ref{Fig:branesW},
for a probe brane in the background of Fig.~\ref{Fig:functionsW}.

\subsection{Summary.}

\begin{figure}
\begin{center}
\begin{picture}(240,400)
\put(0,0){\includegraphics[height=12cm]{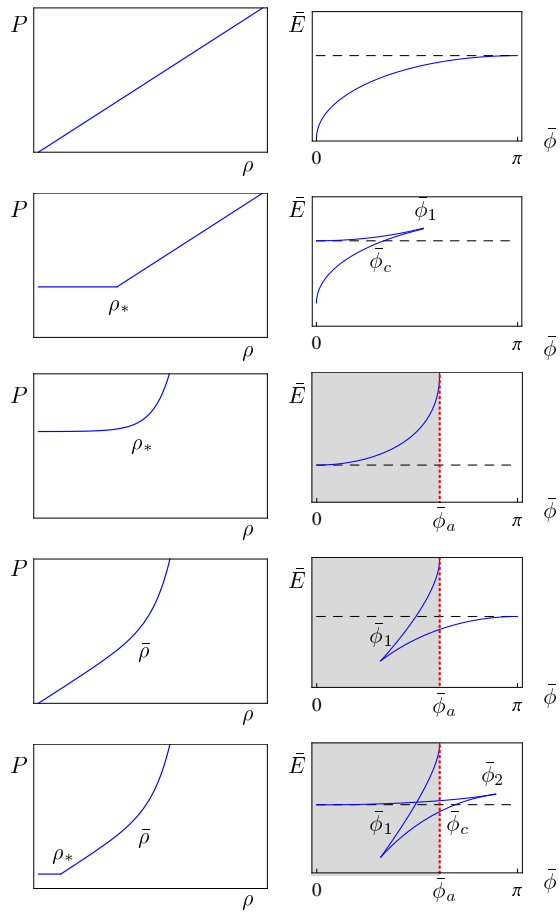}}
\end{picture} 
\caption{Cartoons representing all the possible solutions of the form $P_a$
to the non-Abelian wrapped-D5 system. The left panels show $P$ as a function of $\r$,
while the right panels display, for the corresponding backgrounds,
 the connected (continuous blue lines) configurations for the probe-D7, the disconnected (long-dashed black lines),
 and the short connected configurations (short-dashed red lines).
 The actual plots, rather than these cartoons, can be found either in earlier sections or in~\cite{FPS}.
 The physical configurations are those with lowest $\bar{E}$.
 The shaded region to the left of the short configurations  
 is disconnected from the continuum limit, in the sense that the results 
 do not have an interpretation in the dual field theory.
}
\label{Fig:cartoons}
\end{center}
\end{figure}

We briefly summarise the results for all the 
possible backgrounds controlled by $P_a$, including both those 
discussed here and those that were already examined in~\cite{FPS}.
We present a compendium in Fig.~\ref{Fig:cartoons}.

There are five possible subclasses of supersymmetric solutions to the wrapped-D5 system
that are not fatally singular.
For each class the probe-D7 branes behave differently.
We list them here, following the same order as in the figure.
\begin{itemize}
\item In the CVMN case, for any value of $\bar{\phi}$ there exists only two embeddings.
The connected ones physically represent the chiral-symmetry broken phase. They
 are classically stable (yield the correct concavity in the $(\bar{\phi},\bar{E})$ plane,
and have a negative-definite ${\cal Z}$), and are always the minimum of the energy.
The second configuration allowed is the disconnected one, which is always 
disfavoured. 

\item A third branch appears in backgrounds distinguished by the walking scale $\r_{\ast}$ in the IR, but with the same UV 
behaviour as CVMN. This branch is classically unstable
and a maximum of $\bar{E}$. It is characterised by an endpoint $\hat{\r}_o<\r_{\ast}$.
Connected configurations with $\hat{\r}_o>\r_{\ast}$ are homologous to 
the ones of the CVMN case, but exist only for a range of $\bar{\phi}<\bar{\phi}_1$.
A first-order phase transition emerges with a critical value of $\bar{\phi}_c$
that depends on $\r_{\ast}$. For $\bar{\phi}>\bar{\phi}_c$ the disconnected configuration is dynamically preferred, 
and chiral symmetry is restored. For $\bar{\phi}<\bar{\phi}_c$ the connected solution 
is realised, and the model describes chiral-symmetry breaking.

\item Backgrounds described by the walking scale $\r_{\ast}$ in the IR, 
but for which $P$ grows exponentially in the UV.
There are three branches of solutions. The long connected ones exist only for 
$\bar{\phi}<\bar{\phi}_a=\sqrt{6}\pi/4$, are classically unstable and never a minimum of $\bar{E}$.
The short connected ones, localised near the cutoff, exist for every $\bar{\phi}<\bar{\phi}_a$,
are classically stable, and have divergent (negative) energy. When they exist, they dominate the dynamics.
There is a bulk (first-order) phase transition
at $\bar{\phi}_a$, above which the disconnected configurations yield a 
description of the field theory in the chiral-symmetry restored phase. Below,
the gravity system is not related to the field theory.

\item Backgrounds in which $P$ is approximately linear for $\r<\bar{\r}$, and exponentially growing for $\r>\bar{\r}$.
There are four branches of solutions. Two of them are long and connected, and represent a 
stable branch, which minimizes the energy for $\bar{\phi}>\bar{\phi}_a$,
and an unstable branch that is always subdominant. 
These two branches exist only for $\bar{\phi}>\bar{\phi_1}$, with $\bar{\phi}_1$ a function of $\bar{\r}$,
such that $\bar{\phi}_1<\bar{\phi}_a$.
Disconnected solutions 
are never dominant.
The short connected ones govern the dynamics for $\bar{\phi}<\bar{\phi}_a$,
and a bulk phase transition takes place at $\bar{\phi}_a$. In this instance, the chiral-symmetry broken phase is the physical one connected to the 
field theory.

\item Backgrounds represented by three scales, such that $P$ is approximately constant in
the walking region $\r<\r_{\ast}$, approximately linear for $\r_{\ast}<\r<\bar{\r}$,
and exponentially growing for $\r>\bar{\r}$. There are five branches of  solutions.
There is a first-order bulk phase transition at $\bar{\phi}_a$. Below this value the physics 
is not related to the field theory. Provided the scale $\r_{\ast}\lsim 0.5$, the physical region $\bar{\phi}>\bar{\phi}_a$ 
shows another first-order phase transition at $\bar{\phi}_c>\bar{\phi}_a$, such that for 
$\bar{\phi}_a<\bar{\phi}<\bar{\phi}_c$ the theory 
is in the chiral-symmetry broken phase, and for $\bar{\phi}_c<\bar{\phi}<\pi$ the symmetry is restored. For larger values of $\r_{\ast}$
the physical phase transition disappears, and for any $\bar{\phi}>\bar{\phi}_a$ chiral symmetry is restored.
There are  two turning points at $\bar{\phi}_1$ and $\bar{\phi}_2$, joining different
branches, but both appear in regions that are energetically disfavoured.

\end{itemize}

We now discuss the implications for
physics, in light of the idea that this scenario may be used to 
build models of dynamical EWSB. Given that chiral symmetry is identified with the electroweak gauge group,
it is desirable that the gravity system is in the symmetry-broken phase.

This leaves us with two possibilities. The first is a background with CVMN UV-asymptotics. In this case, $\r_{\ast}$ is unrestricted, but the chiral-symmetry broken 
phase is realised only provided $\hat{\r}_o\gsim \r_{\ast}$. 
It would be interesting to know whether
it is feasible to compute the S-parameter and construct a realistic model.  
For this type of backgrounds,
the calculation of two-point functions are rendered difficult by the non-local behavior of their
divergences in the UV. It is hence not clear at the moment how to define
holographic renormalisation.
On the other hand, it is known that for large values of $\r_{\ast}$ the spectrum of 
fluctuations (glueballs) contains an anomalously light scalar state. This mode is 
probably a dilaton and might coincide with the Higgs particle recently discovered.
This is a compelling scenario that merits further inspection.

The second possibility is an exponential growth for $P$ in the UV.
Necessarily, the model contains two additional scales: $\bar{\r}$,
below which the baryonic VEV becomes important, and a rather small walking scale $\r_{\ast}$,
below which a dimension-six VEV manifests. The shortness of the walking region 
allowed by chiral symmetry breaking means that the effects of walking are probably negligibly small.

\section{Conclusions and further directions.}
\label{Sec:conclusions}

When performing gravity calculations  
for the study of strongly-coupled field theories,
 a regulation procedure is needed. We showed that there are cases in which this may end up revealing that some regions of parameter space 
 are not connected with the field theory. This is analog to bulk transitions in
 lattice gauge theories. The results obtained by naively using the regulation procedure
 are unaffected, as long as the calculations are performed on the right side of the bulk transition.
 All the answers on the wrong side of the transition must be discarded, since they are unrelated to the dual field theory. 
 We exhibited a few examples  and explained them in details. This is a further element in support of the conjectured gauge/gravity dualities.
 
 It would be useful to find a complete (top-down) gravity system  
 that reproduces the phenomenology of dynamical EWSB, including the fact that 
 a Higgs particle has been observed.
 A particularly nice environment is that of the wrapped-D5 system in Type IIB supergravity, wherein a lot of progress has been made.
 Chiral symmetry breaking can be modelled by probing these geometries with D7-branes in a certain special configuration.
 We completed the analysis of all the possible backgrounds of this type. The results are compatible with chiral
 symmetry breaking only in a special subclass, for which precision parameters have not been calculated so far.
 
 This whole program should be repeated in other families of backgrounds,
 starting from the more general class within the Papadopoulos--Tseytlin ansatz~\cite{PT}.
 Some of the observations we made in this paper might provide guidance in this direction.
 It has been found in~\cite{Elander:2014ola} that in the Klebanov--Strassler system there 
 exists (mildly singular) geometries that admit a parametrically light scalar in the spectrum.
 It would be interesting to know whether those backgrounds could be used to model
 dynamical EWSB. The milder UV behaviour of these KS-like solutions
 suggests that it might be possible to calculate the S-parameter, 
 overcoming the difficulties one faces with the wrapped-D5 examples.
 
A completely new feature emerged in the scale-invariant flavoured Abelian backgrounds: the bulk transition is second-order.
 In the proximity of the critical point, gravity effectively describes 
 some (unknown) CFT. It is plausible that similar situations may arise in other instances.
 This is potentially a novel framework to discuss
 new classes of CFT's, and the phenomenology related to the (explicit or spontaneous) breaking
 of their symmetries. For example, this might be an avenue to study the
 four-dimensional dilaton, turning the technical curiosity we uncovered into an approach of practical physics use.

\begin{acknowledgments}
We thank Prem Kumar, Biagio Lucini, Carlos Nu\~nez and Michael Warschawski for useful discussions.
The work of A.~F. was supported by MEC FPA2010-20807-C02-02, by CPAN CSD2007-00042 Consolider-Ingenio 2010 and finally by ERC Starting Grant ``HoloLHC-306605". The work of M.~P. is supported in part by WIMCS and by the STFC grant ST/J000043/1. D.~S. is supported by the STFC Doctoral Training Grant ST/I506037/1.
\end{acknowledgments}



\end{document}